\documentclass{article}
\usepackage{spconf,amsmath,graphicx}
\usepackage{xcolor}
\usepackage{multirow} 
\usepackage{blindtext}
\usepackage{array}
\newcommand{\PreserveBackslash}[1]{\let\temp=\\#1\let\\=\temp}
\newcolumntype{C}[1]{>{\PreserveBackslash\centering}p{#1}}
\newcolumntype{R}[1]{>{\PreserveBackslash\raggedleft}p{#1}}
\newcolumntype{L}[1]{>{\PreserveBackslash\raggedright}p{#1}}

\newcommand{\clotho}{\textit{Clotho}}
\newcommand{\audiocaps}{\textit{AudioCaps}}

\title{Investigations in Audio Captioning: Addressing Vocabulary Imbalance and Evaluating Suitability of Language-Centric Performance Metrics}

\twoauthors
  {Sandeep Kothinti \sthanks{Work done during summer internship at Microsoft Research}}
	{Laboratory for Computational Auditory Perception\\
	Johns Hopkins University, Baltimore, MD, USA\\ sandeep.kothinti@gmail.com}
  {Dimitra Emmanouilidou}
	{Microsoft Research Labs\\
	Redmond, WA, USA\\
 diemmano@microsoft.com}

\begin{document}
\ninept
\maketitle
\begin{abstract}
The analysis, processing, and extraction of meaningful information from sounds all around us is the subject of the broader area of audio analytics. Audio captioning is a recent addition to the domain of audio analytics, a cross-modal translation task that focuses on generating natural descriptions from sound events occurring in an audio stream. In this work, we identify and improve on three main challenges in automated audio captioning: i) data scarcity, ii) imbalance or limitations in the audio captions vocabulary, and iii) the proper performance evaluation metric that can best capture both auditory and semantic characteristics. We find that generally adopted loss functions  can result in an unfair vocabulary imbalance during model training. We propose two audio captioning augmentation methods that enrich the training dataset and the vocabulary size. We further underline the need for in-domain pretraining by exploring the suitability of audio encoders that were previously trained on different audio tasks. Finally, we systematically explore five performance metrics borrowed from the image captioning domain and highlight their limitations for the audio domain.

\end{abstract}


\section{Introduction}
\label{sec:intro}

Recently, technologies for general audio have gained attention, with audio analytics finding uses in security monitoring, content retrieval, and machine monitoring. Several datasets and challenges were developed for audio analytics problems such as audio tagging, audio event detection, etc. Automated audio captioning (AAC) is a relatively new field that builds on various tasks of audio analytics; it is a cross-modal translation task that describes audio contents using natural language.


As with other audio analytics tasks, data-driven models have been a focal point of current research, and represent the backbones of AAC.
Early efforts in AAC focused on collecting large-scale data with high-quality audio captions. AudioCaption \cite{audiocaptions} was one of the early datasets that used videos of web videos and tv programs to collect captions. This work was followed by AudioCaps \cite{audiocaps}, and Clotho \cite{drossos2020clotho}, which collected captions for general audio from the web. While the existing datasets are large, data scarcity is still a problem \cite{mei2022automated}, and any augmentation methods would still benefit data-driven models. And while Mixup \cite{zhang2017mixup} is widely used for augmenting data, it may fall short when it comes to joining captions using natural language. 

When it comes to the available corpora (mentioned above) and the individual data collection studies, each of these efforts followed different data collection paradigms, which lead to datasets with different types of biases. A recent work \cite{martin2021diversity} explored biases from the metadata presented to the annotators and the diversity of captions. However, further biases can also be implicitly imposed - as a result of the structure or syntax of the language itself - and can result in imbalances in the caption vocabulary at the model output. 



The AAC models, at their core, translate a sequence of audio samples into a sequence of words. Following the trend in other sequence models,  initial efforts used recurrent networks \cite{Drossos_2017_waspaa}, replaced later by self-attention-based transformer models \cite{koizumi2020transformer} and a combination of convolutional neural networks (CNN) and transformers \cite{chen2020audio}. The modular nature of CNN-transformer models provides an opportunity to use pretrained models for both the CNNs and the transformer models, which was recently used for fine-tuning the BART model for audio captioning \cite{gontier2021fine}. This also presents the question of what type of pretraining may be best suited for audio captioning. 

Finally, when it comes to model performance, evaluating model-generated captions requires comparing the model output (hypothesis) with reference captions provided by human annotators. In AAC evaluation, while early efforts adopted natural language generation (NLG)~\cite{audiocaptions}, more recent work relies on image captioning metrics that capture the semantic similarity between hypothesis and references~\cite{vedantam2015cider, anderson2016spice}. While these metrics correlate well with human judgments in image captions, their suitability has not been explored for the audio domain. Recently proposed FENSE~\cite{zhou2022can} was found to correlate well with human judgments on audio captions, but since FENSE uses a parametric model, its suitability for AAC is not well explained. 
%
The main contributions of this work are as follows:
\begin{itemize}
    \item We systematically explore evaluation metrics adopted by the image captioning literature for audio captioning. We induce a variety of perturbation errors, similar to perturbation tests performed in NLG metric evaluation~\cite{Sai2021PerturbationCF}, and compare findings side to side with the recently proposed FENSE metric, which was found to better correlate with human judgements.
    \item We explore imbalances in the vocabulary distribution and their effect to model performance, and we further propose a training objective that can help expand the vocabulary size of the model output without sacrificing performance. 
    \item We propose two new approaches for audio caption augmentation, that unlike prior work, use additional  natural language descriptors to join multiple audio and their corresponding captions together.
    \item We investigate three audio encoders that were pretrained on different audio tasks, including audio tagging and multi-modal audio retrieval, and evaluate their suitability for the audio captioning domain.
\end{itemize}




\section{Materials and Methods}
\label{sec:proposed}
\subsection{Datasets}
\label{sec:datasets}
Corpus Clotho v2 (\clotho)~\cite{drossos2020clotho} was used as the primary evaluation dataset for this work (both for training and testing), and  \audiocaps~\cite{audiocaps}  for the augmentation methods described in Section \ref{sec:capaugment}.
\subsubsection{Clotho corpus}
\clotho\ consists of three subsets \textit{development}, \textit{eval} and \textit{validation}, comprising $3,839$, $1,045$ and $1,045$ audio samples, respectively, crawled from freesound database~\cite{font2013freesound}. Given an audio clip, annotators were asked to describe the audio without additional metadata to aid the annotation. Total, five captions were provided by five unique annotators, and further reviewed for grammatical accuracy. The audio was sampled at 16kHz and had a duration range in [15s, 30s]. The vocabulary of the dataset contained a total of $4,300$ words. We chose \clotho\ the primary dataset due to the diversity in vocabulary and number of available captions \cite{martin2021diversity}.


\subsubsection{AudioCaps corpus}
This corpus consists of audio clips from AudioSet~\cite{gemmeke2017audio}, sampled at 16kHz. The \textit{train} set has $\sim$ $50,000$ audio segments, with a single caption only available per audio file. The vocabulary size equals a total of $5,200$ unique words. \audiocaps\ is the largest captioning dataset currently available. We use this corpus as auxilary data for audio captioning augmentation, as described next. We only used samples from \audiocaps\ whose captions contained eight or fewer words.  

\subsection{Caption-based data augmentation}
\label{sec:capaugment}
For audio captioning data augmentation we utilize a joined audio and caption augmentation framework, where the caption augmentation part is governed by the insertion of either temporal or spatial natural language descriptors. We propose two different augmentation methods to enrich the \clotho\ audio and captions: concatenation and mixing. 
For \textit{concatenation}, new audio samples were created by joining audio from \audiocaps\ with audio from \clotho\ either at the beginning or the end with a 50\% chance. For the concatenated samples, the individual captions were combined using conjunction words "and" or "followed by", chosen randomly with a 50\% chance. For \textit{mixing}, the audio files were overlaid using one of the three combination weights, [0.5, 0.5], [0.25, 0.75], or [0.75, 0.25] to create new samples. The shorter samples were zero-padded before mixing. For the mixing strategy, each pair of captions was combined using "and", "in the foreground", or "in the background". These conjunctions were chosen based on the weights applied for mixing the audio. 

\subsection{Proposed model for automated audio captioning}
\label{sec:baseline}
The AAC model used in this work was based on the sequence-to-sequence modeling architecture adopted from task 6A of the DCASE 2022 challenge. In this framework, an audio encoder model processed a 2D time-frequency representation of the audio, such as a spectrogram, to produce a sequence of audio embeddings as the first feature extraction step. These features were passed to a conditional language model, which produced captions in an autoregressive fashion. Details of these steps are described below.

\subsubsection{Audio encoder module} 
Caption generation in the AAC model can strongly depend on the suitability of the audio encoder embeddings. In our experiments, the audio encoder was not fine-tuned during the AAC model training. We explored
three audio encoders: the VGGish \cite{hershey2017cnn} model which was used as the audio encoder in the DCASE 2022 task 6A baseline model. VGGish is a pretrained  \textit{audio tagging} model based on convolutional neural network architecture, with a linear bottleneck layer before the final classification layer. The model was trained on AudioSet to output audio tags from non-overlapping audio segments of duration 0.96s. The model embeddings from the VGGish bottleneck layer were concatenated to be used as the features for the next step.

The second audio encoder was a pretrained CNN14 model from PANNs \cite{kong2020panns}. CNN14 was trained on the AudioSet \textit{audio tagging} task as well, and was  shown to outperform the VGGish model in the tagging performance. The global average pooling layer in CNN14 was replaced with an average-pooling layer with a kernel size of 3. The output of the CNN14 fully connected layer was used as the audio embedding in the same way as the VGGish output. 

The third audio encoder comes from the CLAP \cite{elizalde2022clap} architecture, created for  \textit{audio-retrieval}  and trained on a joint audio and text representation from datasets FSD50k, MACS, AudioCaps. The CLAP audio encoder was based on the CNN14 from PANNs and fine-tuned  for the retrieval task  using audio-text pairs. To disambiguate between the two CNN14s, we referred to CNN14 from the CLAP model as CLAP-CNN. Similar to CNN14, global pooling in CLAP-CNN was replaced by average pooling, and the output of the fully connected layer of the projections in CLAP-CNN was used as the encoder outputs. The hypothesis is that an encoder that has been pretrained on a cross-modal task closer to AAC, in this case CLAP-CNN, may be more appropriate than one pretrained on audio tagging alone.

\subsubsection{Conditional generation}
Following the DCASE 2022 task 6A baseline, the audio embedding vectors were transformed to a 768 dimensional embedding by using an affine transform and were used as an input to a transformer model based on the BART model \cite{lewis2019bart}, a denoising text generator. The encoder part of the transformer processed the input embeddings using six self-attention layers \cite{vaswani2017attention} to produce a transformed embedding vector of 768 dimensions with the same sampling rate as the audio embedding vectors. The decoder model, which had six attention layers, generated a sequence of sub-word units in an auto-regressive process by conditioning the encoder output and an embedding vector of the previously generated sub-word token. The decoder processed the past sub-word embeddings using self-attention and attended to the encoder embedding using cross-attention. The sub-word tokens used in the generation correspond to byte-pair encodings of the words in the training captions. All the layers of the transformer model had 768 dimensional embeddings. The model generated a posterior probability over $50,265$ possible tokens, which could be sampled to produce captions in a sequential process until the model generated an end-of-sentence token. 

\subsubsection{Loss functions}
The baseline AAC model was trained to minimize cross-entropy loss between the generated posterior probabilities and the tokenized ground-truth captions. When multiple captions are available per audio sample, each caption was treated as an independent sample. The cross-entropy loss function weighed the loss for each token equally, and any imbalances in the token frequency were reflected in the model output. We hypothesize that this may be a result of the imbalanced word frequencies found in the training captions, and propose two modifications for exploration. 

First, the use of balanced cross entropy \cite{xie2015holistically}, where a set of prior weights per token were assigned to mitigate the token imbalance, as shown in eq. \ref{ref:eq1}. Here $\omega_c$ indicated class weights, and $\alpha_c$ was the posterior for the ground-truth class. The weights were derived based on the prior counts $p_c$ using a logarithm and a scaling such that the maximum weight was 4. Other transforms or scaling factors did not improve the model performance. \begin{gather}
    \label{ref:eq1}
    \text{balanced cross entropy} = -\omega_c \log(\alpha_c)\\
    \omega_c = -a \log(p_c)
\end{gather}

Another alternative would be to address the token imbalance by imploring the focal loss \cite{lin2017focal}, which was found to generally provide more gains than balanced cross-entropy. Focal loss addresses token imbalance by assigning weights proportional to the prediction errors, which were indirectly caused by the imbalance. The errors were computed based on the prediction probabilities as given in eq. \ref{ref:eq2}.
\begin{gather}
    \label{ref:eq2}
    \text{focal loss} = -(1-\alpha)^\gamma \log(\alpha)
\end{gather}
When using both these loss functions, the training and evaluation procedure was kept the same as the cross-entropy loss.

\subsubsection{Training details}
The baseline model was trained on \textit{Clotho} \textit{development} dataset with cross-entropy loss.  For the loss function comparison between cross-entropy, balanced cross-entropy, and focal loss, the same training and testing procedure was used. For the audio captioning augmentation experiments, the models were trained on both \clotho\ \textit{development} and one of the augmented subsets (\textit{concatenation} or \textit{mixing}).  For each model, the \textit{validation} dataset was used for choosing the best model from the set of models saved during the training. The models were trained for a max of 20 epochs with minibatch size of 8 audio-caption pairs, using AdamW optimizer~\cite{loshchilov2017decoupled}. \textit{Eval} set was used for comparing the model performance.

\subsubsection{Performance Metrics}
AAC model evaluation involves the comparison between natural language descriptions obtained by the audio captioning model (called hypothesis) and those provided by the human annotators (called references).
For this purpose, either natural language generation (NLG) metrics such as BLEU \cite{bleu}, METEOR \cite{banerjee-lavie-2005-meteor} or ROUGE \cite{lin-2004-rouge}, or image captioning metrics such as CIDEr \cite{vedantam2015cider} or SPICE \cite{anderson2016spice} are used in literature. The NLG metrics compare word units such as n-grams (BLEU), the longest subsequence of matching words (ROUGE), or flexible word alignments (METEOR) and do not capture the semantics of the captions. In image captioning evaluations, CIDEr, calculates the cosine similarity of term-frequency inverse document frequency for reference and hypothesis, and SPICE, calculates F-score for graphs with objects, attributes, and relations in the captions. These two metrics were adopted as better alternatives to the NLG metrics. Recently, the FENSE metric \cite{zhou2022can}  was proposed for AAC evaluation, which uses a pretrained sentence similarity model and a fluency penalty to compute score for given reference and hypothesis. While the sentence similarity model used in FENSE was not trained for AAC data, it could predict human judgments better in pairwise comparisons of human and machine-generated captions. In this work, we calculate both FENSE, and FENSE without the fluency penalty (denoted as FENSE*). 

\section{Results}
\label{sec:results}
\subsection{Suitability of performance metrics for perturbation errors}
\label{sec:errors}
\begin{figure}[ht]
    \centering
    \includegraphics[width=\linewidth]{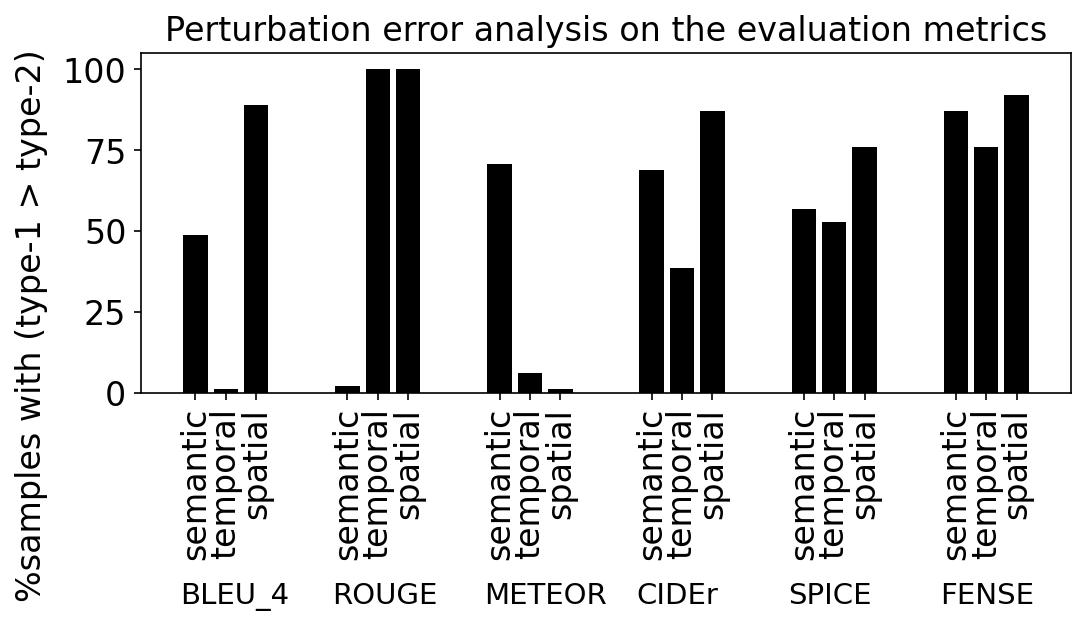}
    \caption{Scores achieved by the five metrics  on the x-axis, for each kind of perturbation error (semantic, temporal, spatial). The y-axis shows the percentage of captions that achieve higher performance score for type-1 errors than for type-2 errors. Metrics with higher values on the y-axis are more suitable for audio captioning.}
    \label{fig:error}
    \vspace{-0.35cm}
\end{figure}

In this section we look at the suitability of the available performance metrics, for AAC. While agreement between the model output (hypothesis) and human judgment (references) provides a way to evaluate the captioning metrics, it does not offer specific reasons why certain metrics are more suitable for AAC. In this work, we utilized three types of perturbation errors that shed light on the suitability of these metrics. In the following scenarios, the type-1 error should be preferred or favored over the type-2 error. In other words, metrics achieving  higher score for type-1 errors than type-2 errors, are more suitable for AAC.
\begin{enumerate}
    \item Semantic order: To test for semantic order, we isolated complex caption sentences comprising two simple sentences combined with the conjunction 'and', from the \textit{development} set.  Type-1 error signifies the swapping (in order) of the two simple sentences within a complex one. For example sentence [A and B] is swapped as [B and A]. Swapping the order of the two simple sentences does not change the semantics and thus should not be penalized.
    In contrast, a type-2 error replaced a verb from one of the simple sentences with a randomly chosen verb while preserving the order of the simple sentences. 
    \item Temporal order: To test for temporal order, we isolated captions containing  temporal keywords 'followed by' or 'and then'. Since auditory phenomena are temporal,  the order of events is often essential. 
    For type-1 errors, these temporal keywords were swapped with simple conjunction 'and'. For type-2 errors, the event order was reversed by swapping the phrases around the temporal keywords. 
    \item Spatial order: Similarly, there is often spatial order in auditory events. Sentences with specific spatial order of events were modified to check if the lack of spatial order ranks higher than the wrong spatial order. For type-1 errors, we swapped spatial keywords (such as 'in the background') with simple conjunctions ('and'). For type-2 errors, the order of the events was reversed by switching the background and foreground events. To test for spatial order, captions containing 'in the background' or 'in the foreground' were used.
\end{enumerate}
For every kind of error, we randomly sampled $1,500$ captions from the \textit{development} set that fit the required structure. For each reference caption,  type-1 and type-2 error captions were compared with the reference to assign a score for each metric. In Fig. \ref{fig:error}, we plot the percentage of captions that achieved higher scores for type-1, than for type-2. The more suitable the metric is, the higher the percentage scores should be.
Across the metrics, FENSE performed consistently better than most other metrics. The NLG metrics failed to generalize across the errors. While ROUGE performed well for temporal and spatial order, it dramatically failed for semantic order, which could be attributed to its sensitivity to the sentence word order. CIDEr and SPICE both showed better consistency than the NLG metrics but perform poorly compared to FENSE. Combining this observation with the accuracy with which FENSE predicts human judgment, FENSE was the best available metric for AAC evaluation. 
For the remainder of the paper, we will drop metrics ROUGE and METEOR.

\subsection{Caption-based data augmentation}
Here we demonstrate the effect of the captioning augmentation methods discussed in Section \ref{sec:capaugment}. Table \ref{tab:aug} rows 1-3 show the performance of the baseline AAC model architecture (trained on \clotho) as compared to the same architecture trained on the two augmented datasets, \textit{concatenation} (concat), and \textit{mixing}. Augmenting with \textit{concatenation} was en par with the baseline in  FENSE, and improved over the baseline in FENSE*, indicating that the model improved in capturing  semantics but was making some syntactical
errors. \textit{Mixing} augmentation seems to hurt  model performance. Note: the conditional generative model was trained almost on the same amount of data; the loss of performance could be attributed to the encoder (VGGish), which was not trained on overlapping auditory events. Importantly, both augmentation methods were able to improve the size of the outputted vocabulary (last column), illustrating the variety added by the augmentation methods. While the augmentation methods showed reserved performance gains on these metrics, they present a novel way to generate more captioning data leveraging  natural language and that can enrich the model vocabulary.  
\begin{table}[ht]
\footnotesize
    \centering
    \begin{tabular}{|p{0.11\linewidth}|p{0.09\linewidth}|p{0.08\linewidth}|p{0.08\linewidth}|p{0.09\linewidth}|p{0.1\linewidth}|p{0.09\linewidth}|}
        \hline
          Method &  BLEU$_4$ & CIDEr & SPICE & FENSE & FENSE* &\#Vocab\\
          \hline 
         \multicolumn{7}{|c|}{Baseline}\\
         \hline
         baseline & 0.154 &0.358 & 0.111 & 0.458 & 0.478 & 298\\
         \hline 
         \multicolumn{7}{|c|}{Data augmentation}\\
         \hline
         Concat & 0.148 & 0.355 & 0.110 & 0.458 & \textbf{0.485} & \textbf{392}\\
         Mixing & 0.130 &  0.308 & 0.112 & 0.449 &0.474 & 352\\
         \hline
         \multicolumn{7}{|c|}{Loss functions}\\
         \hline
         Bal-CE & 0.141 &  0.342 & 0.105 & 0.450 & 0.479 & 332\\
         Focal & 0.128 &  0.323 & 0.104 & 0.457 & 0.481 & \textbf{415}\\
         \hline
    \end{tabular}
    \caption{Performance comparison of the two augmentation methods and the alternative loss functions with the baseline model.}
    \label{tab:aug}
\end{table}
\subsection{Effect of the loss function}
\begin{figure}[ht]
    \centering
    \includegraphics[width=\linewidth]{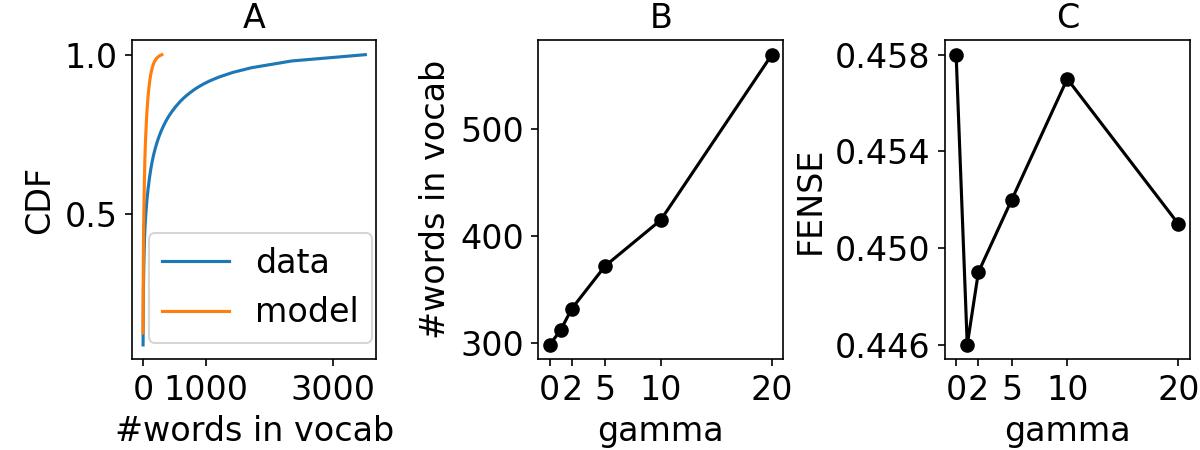}
    \caption{(A) Cumulative distribution of words in the data and model vocabulary with words sorted based on the frequency. (B) Effect of $\gamma$ in focal loss on vocabulary size and (C) on FENSE.}
    \label{fig:fig2}
\vspace{-0.3cm}
\end{figure}

The vocabulary imbalance and its effect on the model are demonstrated in Fig. \ref{fig:fig2}A. The cumulative distribution (CDF) shows that $1,000$ most common words covered about 98\% probability (reminder that \clotho\ has $4,300$ unique words). The baseline model evidently suffers from this, as it notably produces up to 300 unique words (table \ref{tab:aug} first row, last column). 
The last two rows of table \ref{tab:aug} illustrate the increase in model vocabulary without performance sacrificing, when utilizing balanced cross entropy (Bal-CE) or focal loss (focal, $\gamma$=$10$) as the loss function. With a focus on FENSE and FENSE*,
the focal loss seemed to help both the vocabulary size and performed en par with the baseline model. Balanced cross entropy slightly increased the vocabulary size while taking a hit in performance.
Fig. \ref{fig:fig2}B\&C shows the vocabulary size and FENSE values for different values of $\gamma$ in the focal loss. While vocabulary increased with $\gamma$, the FENSE values did not have a monotonic trend. The $\gamma=10$ provided the best FENSE metric among all the values considered. Overall, the focal loss provides a way to increase model vocabulary without requiring additional data or loss on performance. 

\subsection{Effect of different audio encoders}
Invoking audio embeddings from different encoders showed significant effect on the  performance. Table \ref{tab:enc} presents the performance with the different encoders and mAP scores on AudioSet for VGGish (baseline) and CNN14 models. CNN14 improved over the baseline in identifying the audio objects as indicated by the mAP score and this resulted in better performance in the AAC task. CLAP-CNN audio encoder, which was trained on joint audio-text pairs for audio retrieval, outperformed CNN14. This evidence suggested that injecting the natural language information as a retrieval pretraining task can be beneficial for performance and the vocabulary size.
\begin{table}[ht]
\footnotesize
    \centering
    \begin{tabular}{|c|p{0.06\linewidth}|p{0.09\linewidth}|p{0.08\linewidth}|p{0.08\linewidth}|p{0.09\linewidth}|p{0.08\linewidth}|}
        \hline
          encoder & mAP & BLEU$_4$ & CIDEr & SPICE & FENSE & \#Vocab\\
          \hline
         baseline & 0.314  & 0.154 &  0.358 & 0.111 & 0.458 & 298\\
         CNN14 & 0.431   & 0.162 &  0.389 & 0.116 & 0.476 & 305\\
         CLAP-CNN & - &0.157 & 0.400 & 0.118 & 0.479 & 416\\
         \hline
    \end{tabular}
    \caption{Performance comparison of different audio encoders}
    \label{tab:enc}
\end{table}

\section{Conclusion}
\label{sec:conclusion}
In this work, we investigated a multitude of aspects related to automated audio captioning, which we hope could be built upon in future research. Analyzing the effect of induced errors for performance showed that parametric- model metric  FENSE is better suited for Audio Captioning  than simply adopting image captioning metrics. By changing the loss function to suit the prior distribution of the vocabulary, we were able to enrich  the output captions, addressing some of the inherent vocabulary imbalance. We demonstrated that captions could be leveraged to generate new samples that could potentially be used to generate large number of audio-caption pairs. Finally, this work reveals  the need for more complex multi-modal pretraining tasks that also leverage textual description besides audio. 
\section{Acknowledgements}
The authors would like to thank Benjamin M. Elizalde and Soham Deshmukh from Microsoft for providing the CLAP~\cite{elizalde2022clap} encoder.

\bibliographystyle{IEEEbib}
\bibliography{refs.bib}
\label{sec:ref}
\end{document}